\documentclass[a4paper,11pt]{article}
\usepackage[utf8]{inputenc}
\usepackage{color}
\usepackage{xcolor}
\usepackage{amsmath}
\usepackage{amsthm}
\usepackage{adjustbox}
\usepackage{graphicx}
\usepackage{float}
\usepackage{multirow}
\usepackage[mathscr]{eucal}
\usepackage{subfigure}
\usepackage{epstopdf}
\usepackage{caption}
\usepackage{verbatim}
\usepackage{natbib}
\usepackage{colortbl}
 	\definecolor{manatee}{rgb}{0.59, 0.6, 0.67}

\usepackage{lineno,hyperref}

\usepackage{ulem}

\title{Can the hot hand phenomenon be modelled? A Bayesian hidden Markov approach}
\date{October 03, 2023}
 
\author{Gabriel Calvo$^{a}$, Carmen Armero$^{a}$, Luigi Spezia$^{b}$  \\
        \small $^{a}$Department of Statistics and Operations Research, Universitat de València, Burjassot, Spain \\
        \small $^{b}$Biomathematics \& Statistics Scotland, Aberdeen, UK. \\\\
        \small $^{*}$Corresponding author: Gabriel Calvo; \tt{gabriel.calvo@uv.es}
}

\begin{document}

\maketitle
\begin{abstract}
Sports data analytics is a relevant topic in applied statistics that has been growing in importance in recent years. We consider a basketball player or team to have a hot hand when their performance during a game for a certain period of time is better than expected. This phenomenon has generated a great deal of controversy with detractors claiming its non-existence while other authors indicate its evidence. In this work, we present a Bayesian longitudinal hidden Markov model that analyses the hot hand phenomenon in consecutive basketball shots of a team, each of which can be either missed or made. Two possible states (cold or hot) are assumed in the hidden Markov chains of events, and the probability of success for each shot is modelled by considering the corresponding hidden state, random effects associated with the match, and covariates. This model is applied to a real data set, the Miami Heat team in the season 2005-2006 of the USA National Basketball Association. We show that this model is a powerful tool for assessing the overall performance of a team during a game, and, in particular, for quantifying the magnitude of the team streaks in probabilistic terms.
\end{abstract}
\section{Introduction}
The study of success or failure streaks  in sport is a vast issue that has its foundation in the literature of psychology, where it is known as   ``psychological momentum'' \citep{adler1981momentum}. While this subject can be explored in various fields such as politics, finance, or sports, our focus lies solely on the domain of basketball.

The definition of the \textit{hot hand} phenomenon in basketball lacks consensus within the existing literature, although it predominantly revolves around the concept of short-term predictability \citep{green2018hot}. In fact, \citet{avugos2013hot} describe the \textit{hot hand} belief in sports as the conviction that ``following a streak of successful baskets, a player is more likely to continue succeeding with the next shot as well." Similar definitions were adopted by many researchers, including \citet{raab2012hot, miller2018surprised}; and \citet{terner2021modeling}.  However, from our perspective, the most suitable definition, which we solely consider in this work, is that provided by \citet{gilovich1985hot}. The concept suggests that a player has a \textit{hot hand} when their performance during a particular period of time is better than expected on the basis of the player’s overall abilities. Belief in this phenomenon is evident  both in the media and among fans, as well as in many experts from the world of sport.

Generally, in order to demonstrate the existence or absence of this phenomenon, researchers analysed whether the probability of hitting a shot is truly conditioned by a previous streak of successful baskets. This approach was adopted by many researchers, including \citet{gilovich1985hot, raab2012hot, avugos2013hot}; and \citet{miller2018surprised}. The existence of the \textit{hot hand} phenomenon remains a subject of controversy. Some authors, such as \citet{gilovich1985hot}, state that while most fans believe in some sort of correlation among hits and misses in basketball, the {\it hot hand} phenomenon is somewhat of an illusion. On the other hand, other studies contradict Gilovich's findings. For instance, \citet{miller2018surprised} recently argued that there was a clear bias in the analysis conducted by Gilovich.

This paper proposes a novel approach to model the phenomenon known as the \textit{hot hand}. We draw on the works of authors who have explored this behaviour using latent autoregressive structures \citep{mews2023continuous} or  hidden Markov models (HMM) to analyse its occurrence in basketball \citep{sun2004detecting, wetzels2016bayesian, sandri2020} and other sports such as baseball or darts \citep{albert2001streakiness, otting2020hot}. In particular, we propose a Bayesian longitudinal HMM (BLHMM) that analyses the \textit{hot hand} phenomenon in consecutive basketball shots of a team, each of which can be either missed or made. Two possible latent states (\textit{cold} or \textit{hot}) are assumed in the hidden Markov chains of events, and the probability of success for each shot is modelled by considering the corresponding hidden state, random effects associated with the match, and covariates. As a novelty, this model structures each game played by the same team as an observational unit.

The \textit{hot hand} phenomenon has traditionally been analysed from an individual point of view, focusing on assessing the behaviour of a single player. However, our study takes a novel approach by analysing the team. While the majority of the scientific literature on team performance analysis is related to finance \citep{bosch2009knowledge, woolley2009means}, there are also examples of similar analyses in sports \citep{patterson2005influence, thomas2019team}. Furthermore, the existence of a phenomenon known as ``group identity" in sports has been supported by evidence \citep{chen2015competition, heere2011effect}.

We can higlight two important features of our model. First, this model simultaneously assumes dependence between previous shots and the current one through hidden Markov chains. It also assumes that there are moments when the performance of the team is better than others. Secondly, a \textit{hot streak} is defined as a sequence of consecutive moments in which the team is in the \textit{hot} state. Conversely, a \textit{cold streak} is a sequence of consecutive shots in which the team is in the \textit{cold} state.

The structure of the paper is as follows. Section 2 introduces the BLHMM   in terms of a sampling model based on two connected processes, one latent (or hidden) and the other one observed,  and a prior distribution for all of the   unknown elements of the model. Section 3 deals with the posterior distribution for some characteristics of the basketball team performance such as  transition probabilities, occupancy and sojourn times, and probabilities of making a basket. 
 Section 4 applies the Bayesian model to a real data set from the Miami Heat team in the season 2005-2006 of the National Basketball Association (NBA).  The paper concludes with a discussion, including potential directions for future research.

\section{Bayesian longitudinal hidden Markov modelling} \label{sec:model}

 In the Bayesian framework, probabilities are always conditional and will therefore appear as such in their corresponding descriptions. This situation increases the complexity of the notation but clarifies much better the theoretical framework of Bayesian inference. 
 
 We propose a BLHMM for analysing the shooting performance of a basketball team in a season or a group of $N$ games as a joint probabilistic model $f(\boldsymbol{Y}, \boldsymbol{Z}, \boldsymbol{\theta}, \boldsymbol{\psi})$ for the observed process $\boldsymbol Y$,  the hidden process  $\boldsymbol Z$, the random effects $\boldsymbol \psi$ associated with both processes,  and the parameters and hyperparameters $\boldsymbol \theta$.  This joint distribution can be factorised as

\begin{equation}
  f(\boldsymbol{Y}, \boldsymbol{Z}, \boldsymbol{\theta}, \boldsymbol{\psi}) = \underbrace{ f(\boldsymbol{Y} \mid \boldsymbol{Z}, \boldsymbol{\theta}, \boldsymbol{\psi}) }_{\text{Observable process}} \ 
  \underbrace{ f(\boldsymbol{Z} \mid \boldsymbol{\theta}, \boldsymbol{\psi})}_{\text{Hidden process}}\  f(\boldsymbol{\psi}|\boldsymbol \theta) \ \pi(\boldsymbol{\theta}). \label{joint}
\end{equation}

We assume a general framework of conditional i.i.d. among the different games of the season ($i=1,\ldots,N$), so that
\begin{equation}
  f(\boldsymbol{Y}, \boldsymbol{Z}, \boldsymbol{\theta}, \boldsymbol{\psi}) =   \big(\prod_{i=1}^{N} \, f(\boldsymbol{Y}_i \mid \boldsymbol{Z}_i, \boldsymbol{\theta}, \boldsymbol{\psi})    \ 
  f(\boldsymbol{Z}_i \mid \boldsymbol{\theta}, \boldsymbol{\psi}) \big)\  f(\boldsymbol{\psi}|\boldsymbol \theta) \ \pi(\boldsymbol{\theta}). \label{joint2}
\end{equation}

We describe below the different probabilistic elements in (\ref{joint2}), with the first two expressed  in terms of a generic game $i$.\\

 \noindent \textbf{The hidden process.} This is a hidden Markov chain  (HMC) that describes the {\it hot hand} situation of shots at the basket of the team  in game $i$. The transition probabilities of the HMC are  modeled using logistic mixed regression models. \vspace*{-0.4cm}\\

    Let  $\{Z_{in}, n = 1,\dots,M_i\}$ be a   HMC which  represents the state, {\it cold }($C$) or {\it hot} ($H$), of the team  in shooting $n$ at the basket in game $i$, where $M_i$ is the number of shots in the game $i$. The transition probability matrix of the HMC is expressed as
    
    \begin{equation}
P_i  = \begin{pmatrix}
 p_i^{(CC)} & p_i^{(CH)} \\ 
p_i^{(HC)} & p_i^{(HH)}
\end{pmatrix},\label{transition1}
\end{equation}

\noindent  where $p_i^{(CH)}=P(Z_{i,n+1}=H \mid Z_{in}=C, \boldsymbol \theta, \boldsymbol \psi)$ is the conditional (on $\boldsymbol \psi$ and $\boldsymbol \theta$) transition probability from the {\it cold }  $C$
to the {\it hot} state $H$ in the game $i$,    $p_{i}^{(CC)}=1-p_i^{(CH)}$,
$p_i^{(HC)}=P(Z_{i,n+1}=C \mid Z_{in}=H, \boldsymbol \theta, \boldsymbol \psi)$, and $p_{i}^{(HH)}=1-p_i^{(HC)}$. Thus, the symbol $\boldsymbol{Z}_i$  of Formula (\ref{joint2}) is the random vector $(Z_{i1}, \ldots, Z_{iM_{i}})$.

The complete specification of the Markov chain needs to set a probability distribution for the initial state of the chain  $\boldsymbol{\delta}_i=(\delta_{i}^{(C)}, \delta_{i}^{(H)}) =(P(Z_{i1}=C \mid \boldsymbol \theta, \boldsymbol \psi),\, P(Z_{i1}=H \mid \boldsymbol \theta, \boldsymbol \psi))$. We know that, from a theoretical standpoint, the values of the discrete latent states are natural numbers and not letters. However, for the sake of readability, we chose to represent the values of the variables in the chain with letters, $C$ and $H$.

We assume a logistic mixed regression model for the transition probabilities as follows:
\begin{align*}
\text{logit}(p_i^{(CH)}  \mid \boldsymbol \theta, \boldsymbol \psi) &= \boldsymbol{X}_i \boldsymbol{\beta}^{(CH)} + b_{i}^{(CH)},\\
\text{logit}(p_i^{(HC)} \mid \boldsymbol \theta, \boldsymbol \psi)&= \boldsymbol{X}_i \boldsymbol{\beta}^{(HC)} + b_{i}^{(HC)},
\end{align*}
 \noindent where  $\boldsymbol{X}_i$ is a vector of baseline covariates, and   $\boldsymbol{\beta}^{(CH)}$ and $ \boldsymbol{\beta}^{(HC)}$ are regression coefficient vectors  associated   with transitions from $C$ to $H$   and from $H$ to $C$, respectively.  The random effects associated with these transition probabilities  within the game  are  $\boldsymbol b_i=(b_{i}^{(CH)}, b_{i}^{(HC)})$,  which are assumed to follow a conditional multivariate normal distribution as
 $(\boldsymbol b_{i}|\boldsymbol \theta) = \text{N}(\boldsymbol{0},\Sigma_{b})$, where $\Sigma_{b}$ is a variance-covariance matrix. Note that  $\boldsymbol{\beta}=(\boldsymbol{\beta}^{CH)}, \boldsymbol{\beta}^{(HC)})$ and $\Sigma_b$ are parameters and hyperparameter included in the generic $\boldsymbol \theta$ vector. The same applies to the random effects $b_{i}$'s as elements of $\boldsymbol \psi$. \vspace*{-0.1cm}\\

\noindent \textbf{The observed process.} A Bernoulli longitudinal model is used to   assess the success or failure of shots at the basket in each game $i$,  with  probability depending on the state ({\it cold }or {\it hot}) of the team. These state-dependent probabilities are modelled through  logistic mixed regression models.\vspace*{-0.4cm}\\

Let the random variable $Y_{in}$ be  the success (1) or failure (0) of shot $n$ in game $i$ of the team. Thus, the symbol $\boldsymbol{Y}_i$ of Formula \ref{joint} is the random vector $(Y_{i1}, \ldots, Y_{iM_{i}})$. Each variable in ${Y}_{in}$    is a Bernoulli whose probability depends on the  latent state $Z_{in}$ as follows, 
\begin{align*}
(Y_{in}\mid \boldsymbol{\theta},  \boldsymbol \psi, Z_{in}=C)& \sim \text{Bern}(\gamma_{in}^{(C)}),\\
(Y_{in}\mid\boldsymbol{\theta},  \boldsymbol \psi, Z_{in}=H)& \sim \text{Bern}(\gamma_{in}^{(H)}),
\end{align*}

\noindent 
where $\gamma_{in}^{(H)}$ and $\gamma_{in}^{(C)}$ are the probability parameter associated with the 
{\it hot} and {\it cold }states, respectively. The probabilities vectors $\boldsymbol{\gamma}_{i}^{(C)} = (\gamma_{i1}^{(C)}, \dots, \gamma_{iM_i}^{(C)})$ and $\boldsymbol{\gamma}_{i}^{(H)} = (\gamma_{i1}^{(H)}, \dots, \gamma_{iM_i}^{(H)})$  can be modelled through a logistic mixed regression model  expressed as
\begin{align*}
\text{logit}(\gamma_{i}^{(C)} \mid \boldsymbol \theta, \boldsymbol \psi) &= \mathcal{X}_{i} \boldsymbol{\alpha}^{(C)} + \Xi_{i} \boldsymbol{a}_i^{(C)},\\
\text{logit}( \gamma_{i}^{(H)} \mid \boldsymbol \theta, \boldsymbol \psi) &= \mathcal{X}_{i} \boldsymbol{\alpha}^{(H)} + \Xi_{i} \boldsymbol{a}_i^{(H)},
\end{align*}

\noindent where $\mathcal{X}_i$ and $\Xi_i$ are the design matrices for the fixed effects $\boldsymbol{\alpha}= (\boldsymbol{\alpha}^{(C)}, \boldsymbol{\alpha}^{(H)})'$  and the random effects  $\boldsymbol{a}_i= (\boldsymbol{a}_i^{(C)}, \boldsymbol{a}_i^{(H)})'$, respectively. Matrix $\mathcal{X}_i$ includes the covariates associated with the success of shooting, for instance the distance to the basket or the type of shot.  Random effects 
$\boldsymbol{a}_i$ are  conditionally independent and assumed to follow a conditional multivariate normal distribution  $(\boldsymbol{a}_i|\Sigma_a) \sim \text{N}(\boldsymbol{0},\Sigma_a)$. Note that  parameters $\boldsymbol \alpha$ and hyperparameters in $\Sigma_a$ are part of $\boldsymbol \theta$ (along with $\boldsymbol \beta$, $\Sigma_b$, and $\boldsymbol \delta= (\boldsymbol{\delta}_{1},\ldots,\boldsymbol{\delta}_{N})$), whereas  the  $\boldsymbol a_i$'s of $\boldsymbol \psi$ (along with $\boldsymbol b=(\boldsymbol{b}_{1},  \ldots, \boldsymbol{b}_{N})$). 
Therefore, $\boldsymbol \theta=(\boldsymbol \alpha, \boldsymbol \beta, \Sigma_{a}, \Sigma_{b},\boldsymbol \delta)$ and $\boldsymbol \psi=(\boldsymbol a, \boldsymbol b)$, where $\boldsymbol a=(\boldsymbol a_{1},\ldots,\boldsymbol a_{N})$.

The Bayesian  model is completed  with the specification of  a prior distribution $\pi(\boldsymbol \theta)$ over the parameters and hyperparameters of the model.

\section{Basketball team performance }\label{sec:team}

Bayes' theorem combines the information provided by the prior distribution and the likelihood function to obtain the posterior distribution $\pi(\boldsymbol{\theta}, \boldsymbol{\psi} \mid \mathcal D)$, where $\mathcal D$ stands for the data  and includes the observations $\boldsymbol{y}_i$,(i.e., the realisations of the process $\boldsymbol Y_i$ of Formula \ref{joint}, that is, the baskets made and missed of the shots of the games)   and  a few  covariates. This posterior contains all information of the behaviour of the model and allows direct knowledge of the performance of the team both at a general level (such as the coefficients of the logistic regressions associated with either the hidden transition probabilities or the state dependent probabilities of making a basket) and at a specific level of the different games in the analysed championship. At the specific level, the posterior distribution provides knowledge on the random effects associated with a particular game and team status on transition probabilities or probabilities of making a basket. That information is very valuable but   does not provide direct knowledge on some of the indicators of the team performance that may be important to better understand its strengths and weaknesses. This is the case of transition probabilities,  probability of making a basket, probability of being in a {\it cold }or a {\it hot} state, occupancy times, first-passage times, etc. Since the stochastic behaviours of these quantities depends on $(\boldsymbol \theta, \boldsymbol \psi)$, we can apply the Bayesian paradigm and compute the posterior distribution of each of them from the   posterior  distribution  $\pi(\boldsymbol{\theta}, \boldsymbol{\psi} \mid \mathcal D)$. We discuss these inferences below.

\subsection{Transition probabilities}\label{subsec:transition}

 For the sake of space, we will only develop the results of the inference for the transition probabilities of the HMC defined in (\ref{transition1}) from the {\it cold }to the {\it hot }state, i. e. $p_i^{(CH)}$.  The same strategy can be applied to the other transition probabilities  of the chain.  Therefore,
\begin{equation}
    P(Z_{i,n+1}=H \mid Z_{in}=C, \boldsymbol \theta, \boldsymbol \psi)= \frac{\mbox{exp}\{\boldsymbol{X}_i \boldsymbol{\beta}^{(CH)} + b_{i}^{(CH)}\}}{1+\mbox{exp}\{\boldsymbol{X}_i \boldsymbol{\beta}^{(CH)} + b_{i}^{(CH)}\}}. \label{transition2}
    \end{equation}
    
\noindent Since this transition probability depends on   $\boldsymbol \theta$  and $\boldsymbol \psi$, we can   approximate its posterior distribution,

\begin{equation}
    \pi( P(Z_{i,n+1}=H \mid Z_{in}=C, \boldsymbol \theta, \boldsymbol \psi) \mid \mathcal D),
\end{equation}

 \noindent  through a simulated sample from the posterior $\pi(\boldsymbol{\theta}, \boldsymbol{\psi} \mid \mathcal D)$. Thus, we can compute   the relevant characteristics of that distribution (mean, standard deviation, credible intervals, etc).  If we also want to evaluate these probabilities in a general way, without distinguishing between games, we can obtain the marginal transition probability
\begin{equation}
    P(Z_{i\,n+1}=H \mid Z_{in}=C, \boldsymbol \theta)= \int P(Z_{i\,n+1}=H \mid Z_{in}=C, \boldsymbol \theta, \boldsymbol \psi)\, f(\boldsymbol \psi \mid \boldsymbol \theta) \, \mbox{d}\boldsymbol \psi, 
\end{equation}

\noindent which is independent of the game and depends  on $\boldsymbol \theta$ only. Consequently,  its posterior distribution, 
\begin{equation}\label{eq:int_re}
    \pi( P(Z_{i,n+1}=H \mid Z_{in}=C, \boldsymbol \theta) \mid \mathcal D),
\end{equation}
\noindent can be approximated using the conditional posterior distribution $\pi(\boldsymbol{\theta}  \mid \mathcal D)$ which is easily obtained through a simulated sample from $\pi(\boldsymbol{\theta}, \boldsymbol \psi  \mid \mathcal D)$.

\subsection{t-step transition probabilities and  stationary distribution}

The $t$-step transition probabilities of the chain for game $i$ are defined as

\begin{align*}
    p_i^{(CH)}(t)= & P(Z_{i\,n+t}=H \mid Z_{in}=C, \boldsymbol \theta, \boldsymbol \psi), \,\, p_i^{(CC)}(t)=1-p_i^{(CH)}(t),\\
    p_i^{(HC)}(t)= & P(Z_{i\,n+t}=C \mid Z_{in}=H, \boldsymbol \theta, \boldsymbol \psi), \,\, p_i^{(HH)}(t)=1-p_i^{(CH)}(t).
\end{align*}

\noindent From the Markov chain theory (e.g., \cite{kulkarni2016modeling}) we know  that the matrix that collects these probabilities, $P_i^{(t)}$, can be calculated from the transition probability matrix $P_i$ in (\ref{transition1}) as 
\begin{equation*}
     \begin{split}
{P_i}^{(t)}    & =  \frac{1}{p_{i}^{(CH)}+p_i^{(HC)}} \, 
 \left[ \begin{array}{cc}   p_{i}^{(HC)} & p_{i}^{(CH)}\\  p_{i}^{(HC)} & p_{i}^{(CH)}   \end{array} \right] \\ \\
 & +  \frac{(1-p_{i}^{(CH)}-p_i^{(HC)})^{t}}{ p_{i}^{(CH)}+p_i^{(HC)}} \, 
 \left[ \begin{array}{cc}   p_{i}^{(CH)} & -p_{i}^{(CH)}\\  -p_{i}^{(HC)} & p_{i}^{(HC)}   \end{array} \right].
 \end{split}
\end{equation*}

\noindent Since the HMC for game $i$  has a finite state space, and it is irreducible and aperiodic, its stationary distribution will be the limit distribution,  $\boldsymbol{\Delta}_i= (\Delta_i^{(C)}, \Delta_i^{(H)})$, which turns out to be
\begin{equation}
\begin{split}
    \Delta_i^{(C)} &= \frac{p_{i}^{(HC)}}{p_{i}^{(CH)}+p_i^{(HC)}},\,\,\,\,
    \Delta_i^{(H)} = \frac{p_{i}^{(CH)}}{p_{i}^{(CH)}+p_i^{(HC)}}.
\end{split}    
\end{equation} 

\noindent Note that the vector $\boldsymbol{\Delta}_i$ is also the left eigenvector of the transition probability matrix $P_i$, associated with the eigenvalue one.

The Bayesian framework allows us to calculate the  posterior distribution of each of the $t$-step transition probabilities of the chain for each specific game as well as for a generic game by means of the posterior distribution of the conditional marginal distribution as explained in Subsection 3.1.

\subsection{Occupancy times}
Occupancy times in a chain  refers to the expected number of visits the process makes to each state of the chain in a given number of transitions \citep{kulkarni2016modeling}. This  is an indicator   of the frequency with which the chain visits the different states of the process. In particular, for each game $i$ we represent by 
$m_i^{(jk)}(t)=\mbox{E}(V_i^{(jk)}(t) \mid \boldsymbol \theta, \boldsymbol \psi)$   the conditional (on 
$(\boldsymbol \theta, \boldsymbol \psi)$) expected value  of the number of visits $V_i^{(jk)}(t)$ to state $k$    from state $j$ in the first  $t$ transitions of the chain.  In the case of our HMC, these conditional expectations are expressed \citep{kulkarni2016modeling} in terms of the transition probabilities as follows
 
\begin{equation}
    \begin{split}
 \begin{bmatrix} m_i^{(CC)}(t)  &  m_i^{(CH)}(t) \\ m_i^{(HC)}(t)  & m_i^{(HH)}(t)   \end{bmatrix} &   = 
\frac{t+1}{p_{i}^{(CH)}+p_i^{(HC)}} \, 
 \left[ \begin{array}{cc}   p_{i}^{(HC)} & p_{i}^{(CH)}\\  p_{i}^{(HC)} & p_{i}^{(CH)}   \end{array} \right] \\ \\  
    & \hspace*{-1.5cm}  +\frac{1-(p_{i}^{(HC)}+p_i^{(CH)}-1)^{(t+1)}}{(p_{i}^{(CH)}+p_i^{(HC)})^2} \, \,
 \left[ \begin{array}{cc}   p_{i}^{(CH)} & -p_{i}^{(CH)}\\  -p_{i}^{(HC)} & p_{i}^{(HC)}   \end{array} \right] 
 \end{split}
\end{equation}

 For inferential purposes, we should proceed in the same way as we did when hadling the transition probabilities. We can compute the posterior distribution of each occupancy time. This give us information on each game. In a general way, we can integrate out the random effects associated with the individual games and obtain posterior information  of the number of visits among a given number of transitions in a generic game. 

\subsection{Sojourn times}

The sojourn time of the basketball team in state $j$, where $j \in \{C,H\}$, during game $i$, denoted by $\tau_{i}^{(j)}$, is defined as the number of shots required for the team to transition out of state $j$ for the first time, i.e.
\begin{align*}
    P(\tau_{i}^{(j)}=t \mid \boldsymbol \theta, \boldsymbol \psi)& = P(Z_{i\,n+t} \neq j, Z_{i\, t^*} = j, t^*=n+1, \ldots, n+t \mid Z_{in} = j, \boldsymbol \theta, \boldsymbol \psi)\\
    & =[p_i^{(jj)}]^{t}\,(1-p_{i}^{(jj)}), \,\, t=0,1,\ldots
\end{align*}

\noindent Therefore, the conditional distribution of the sojourn time in state $C$ ($H$) of the game $i$ will be a geometric distribution of   parameter  $p_{i}^{(CC)}$ ($p_{i}^{(HH)}$).  Consequently, we can 
compute the posterior distribution for $\tau_{i}^{(C)}$ and $\tau_{i}^{(H)}$
\begin{equation}
    \begin{split}
       &\pi((\tau_{i}^{(C)}=n \mid \boldsymbol \theta, \boldsymbol \psi) \mid \mathcal{D}),\\
       &\pi((\tau_{i}^{(H)}=n\mid \boldsymbol \theta, \boldsymbol \psi) \mid \mathcal{D}),
    \end{split}
\end{equation}
 \noindent or its marginal posterior distribution integrating out the random effects as above.
 
\subsection{Probabilities of making a basket}

The procedure for inferring the probability of making the $n$th shot of game $i$ when the team is in a {\it hot }or {\it cold }state is similar to the one we have developed for calculating the  posterior  distribution of transition probabilities in subsection (\ref{subsec:transition}). Thus, the relevant conditional posterior distributions are
\begin{equation}
    \begin{split}
        \pi(P(Y_{in}=1 \mid Z_{in}=C, \boldsymbol \theta, \boldsymbol \psi) \mid \mathcal D) \nonumber\\
\pi(P(Y_{in}=1 \mid Z_{in}=H,   \boldsymbol \theta, \boldsymbol \psi) \mid \mathcal D).
    \end{split}
\end{equation}

 If interested in a general performance of the team, independently of the game, we can also marginalise the random effects associated with the games, compute the conditional marginal distributions  $P(Y_{in}=1 \mid Z_{in}=C, \boldsymbol \theta, \mathcal D)$ and  $P(Y_{in}=1 \mid Z_{in}=H, \boldsymbol \theta,\mathcal D)$, and its subsequent posterior distributions
 \begin{equation}
    \begin{split}
        \pi(P(Y_{in}=1 \mid Z_{in}=C, \boldsymbol \theta ) \mid \mathcal D) \nonumber\\
\pi(P(Y_{in}=1 \mid Z_{in}=H,   \boldsymbol \theta ) \mid \mathcal D).
    \end{split}
\end{equation}
 
In addition, we can compute the posterior distribution of the probability of making the $n$th shot of game $i$ when the state of the team  is unknown, i. e. $\pi(P(Y_{in}=1 \mid  \boldsymbol \theta ) \mid \mathcal D)$. This quantity can be simulated from the posterior distribution $\pi( \boldsymbol \theta, \boldsymbol \psi \mid \mathcal D)$ and 
\begin{align*}
      P(Y_{in}=1 \mid   \boldsymbol \theta, \boldsymbol \psi) &= P(Y_{in}=1 \mid Z_{in}=C,  \boldsymbol \theta, \boldsymbol \psi)P(Z_{in}=C \mid    \boldsymbol \theta, \boldsymbol \psi)  \\
      &+ P(Y_{in}=1 \mid Z_{in}=H, \boldsymbol \theta, \boldsymbol \psi) 
      P(Z_{in}=H \mid  \boldsymbol \theta, \boldsymbol \psi),
\end{align*}
where 
\begin{align*}
      P(Z_{in}=C \mid   \boldsymbol \theta, \boldsymbol \psi) &= P(Z_{in}=C \mid  Z_{i1}=C, \boldsymbol \theta, \boldsymbol \psi)\,P(Z_{i1}=C \mid    \boldsymbol \theta, \boldsymbol \psi)  \\
      &+ P(Z_{in}=C \mid  Z_{i1}=H, \boldsymbol \theta, \boldsymbol \psi)\,P(Z_{i1}=H \mid    \boldsymbol \theta, \boldsymbol \psi).
  \end{align*}

\section{Miami Heat shooting in the 2005-2006 NBA season}

On 20 June 2006, the Miami Heat team defeated the Dallas Mavericks in the finals and won their first NBA championship. The analysis of the information play-by-play provided by the \cite{pbp} about the team is very interesting and gives many clues on their successful season. Specifically, we are interested in analysing  their shooting performance.
We consider all of  the {\it field shots} and {\it free throws} of all games of the championship,   a total of $11\hspace*{0.35mm}042$ shots   in 105 games,   $5\hspace*{0.35mm}922$ of which were baskets made. For each of these shots the distance in feet from the shooting position to the basket, the game, and the sequential order per game in which each shot was taken were taken into account. 

\begin{figure}[ht]
\begin{center}
\includegraphics[width= 12cm]{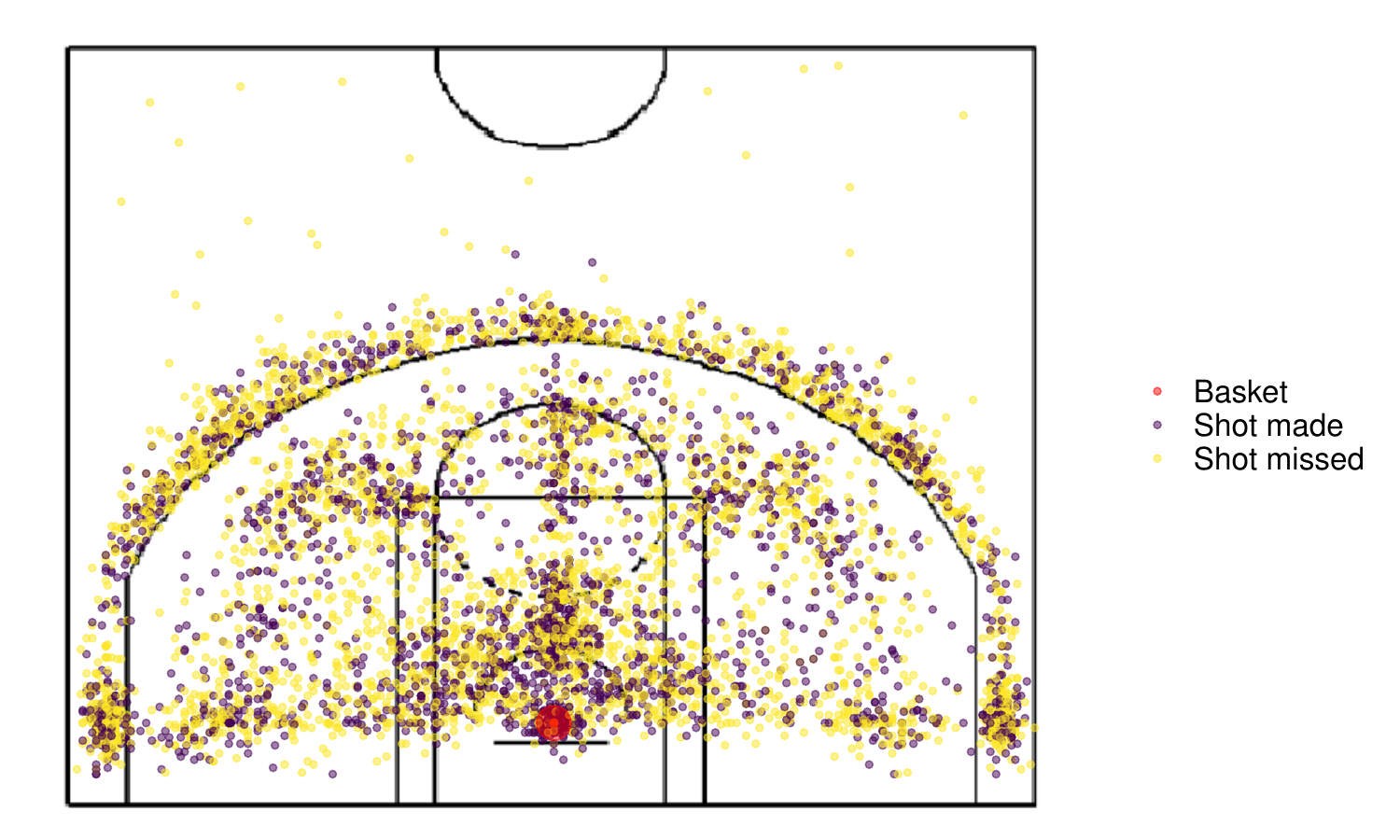} 
\caption{Shot chart of the Miami Heat during the 2005-2006 season, showing the location of shots on the court, and indicating  the shot results and the exact position of the basket. } \label{fig:chart}
\end{center}
\end{figure}

Figure \ref{fig:chart} is a shot chart \citep{zuccolotto2020basketball}  of the Miami Heat's {\it field shots}  during the 2005-2006 NBA season. It can be clearly seen that most shots are taken just beyond the {\it three-point line} or just below the basket. On the other hand, as it might be expected, most of the shots the team took from further away were missed.

\subsection{Modelling the shooting performance of the Miami Heat team}

We apply the Bayesian model  presented in Section (\ref{sec:model}). The sampling model is also defined in terms of the two sub-processes, one defined in terms of an HMC that accounts for the {\it hot }and {\it cold }states, and the other in terms of an observed  Bernoulli longitudinal variable that assesses the success or failure of each shot in relation to the {\it hot }or {\it cold }state of the chain.

The transition probabilities of the HMC $\{Z_{in}, n = 1,\dots,M_i\}$ defined for each game $i$, $i=1,\ldots, N=105$, are expressed through the logistic mixed regression models 
\begin{align*}
\text{logit}(p_i^{(CH)}  \mid \boldsymbol \theta, \boldsymbol \psi) &=  {\beta}_{CH} + b_{i}^{(CH)},\\
\text{logit}(p_i^{(HC)} \mid \boldsymbol \theta, \boldsymbol \psi)&= {\beta}_{HC} + b_{i}^{(HC)},
\end{align*}

\noindent which express each of the two relevant probabilities, $p_i^{(CH)}$ and $p_i^{(HC)}$,   in terms of an intercept (indicating the average logit of the transition probabilities), common to all games and a specific random effect, $b_{i}^{(CH)}$ and
$b_{i}^{(HC)}$, associated with the game.  These random effects are mutually independent and conditionally normally distributed as $(b_{i}^{(CH)}| \sigma_{CH})\sim \text{N} (0,  \sigma_{CH}^2)$ and $(b_{i}^{(HC)}|  \sigma_{HC})\sim \text{N} (0,  \sigma_{HC}^2)$.
The initial probability vector $\boldsymbol{\delta}=(\delta_{C}, \delta_{H})$ is also considered. 

For the observed part of the sampling model, the probabilities  $\gamma_{in}^{(H)}$ and $\gamma_{in}^{(C)}$   of making a basket on shot $n$ in game $i$ when the team is respectively in the {\it hot }or the {\it cold }state are described as 
\begin{align*}
\text{logit}(\gamma_{in}^{(C)}) &= \alpha_{C} + \alpha_{d}X_{in} + \alpha_{FT} I_{FT}(in) + a_i,\\
\text{logit}(\gamma_{in}^{(H)}) &= \alpha_{H} + \alpha_{d}X_{in} + \alpha_{FT} I_{FT}(in) + a_i,
\end{align*}

\noindent where $\alpha_{C}$ and $\alpha_{H}$ are common intercepts for the probability associated with the {\it cold }and the {\it hot }state respectively, $\alpha_d$ and $\alpha_{FT}$ are the regression coefficients associated with covariates $X_{in}$ and $I_{FT}(in)$  that respectively  describe the 
distance from the basket in the $n$-th shot of the game $i$ and   an indicator variable that is 1 when the $n$-th shot of game $i$ is a {\it free throw} and zero otherwise. Random effects $a_i$ are assumed normally distributed, i. e. $(a_i|\sigma_a) \sim \text{N}(0,\sigma_a^2)$, and  conditional independent given $\sigma_a$, for any $i$.

In order to complete the specification of the BLHMM model we need to elicit a prior distribution for the parameters and hyperpameters of the model. We assume prior independence within a minimally informative prior scenario. In particular, we select a beta distribution for the initial probability $\pi(\delta_{C})= \text{Be}(1, 1)$, and uniform distributions for the standard deviation parameters $\pi (\sigma_a)=\pi (\sigma_{CH})=\pi (\sigma_{HC})=\text{U}(0, 10)$. For the two regression coefficients, $\alpha_{d}$ and $\alpha_{FT}$, associated with the covariates   a wide normal distribution is selected, $\pi (\beta_{CH})=\pi (\beta_{HC})=\pi (\alpha_{d})=\pi (\alpha_{FT})=\text{N}(0, 10^2)$. Finally, to avoid the problem of identifiability because of the label switching
issue (see among others \citealp{mcculloch1994statistical, fruhwirth2001markov, spezia2009reversible}), we include the following restriction in the $\alpha_C$'s and  $\alpha_H$'s prior distributions
\begin{equation*}
\begin{split}
\pi(\alpha_{C})=\pi(&\alpha_{H})= \text{N}(0, 10^2),\,\,
\alpha_{C} \leq  \alpha_{H}.
\end{split}
\end{equation*}

\subsection{Posterior distribution}

The complexity of the BLHMM model makes the posterior distribution analytically intractable. We approximated it by means of  Markov chain Monte Carlo (MCMC) sampling methods \citep{tanner2012tools} via the JAGS software \citep{plummer2003jags}. Three parallel chains were run for $30\hspace*{.35mm}000$ iterations each after burn-ins of $30\hspace*{.35mm}000$ iterations. In addition, based on the estimated autocorrelation in the sample, and in order to reduce it, the chains were also thinned at every 30th iteration. Moreover, the full analysis, performed by an R code (R version 4.0.5), and the data are available as supplementary material at \url{https://github.com/gcalvobayarri/hot_hand_model.git}.

	 \begin{table}[H]
		\centering
		\caption{Posterior summaries (mean, standard deviation, 95\% credible interval) and diagnostic statistic $\hat{R}$ for the parameters and hyperparameters of the Miami Heat shooting performance {\it hot hand} model.}\label{tab:post}
		\vspace{0.25cm}
		\begin{tabular}{ccccccc}
			\noalign{\hrule height 1pt}
			\multicolumn{1}{l}{Submodel} & \multicolumn{1}{l}{} & \multicolumn{1}{c}{mean}    & \multicolumn{1}{c}{sd} & $q_{0.025}$ & $q_{0.975}$ & $\hat{R}$ \\ \noalign{\hrule height 1pt}
			\multicolumn{1}{l}{Latent} & \multicolumn{1}{c}{$\beta_{CH}$}  & \multicolumn{1}{r}{$-0.49$} & \multicolumn{1}{r}{$0.05$} & \multicolumn{1}{r}{$-0.58$} & \multicolumn{1}{r}{$-0.39$} & $1.001$\\
			 & \multicolumn{1}{c}{$\beta_{HC}$}     & \multicolumn{1}{r}{$0.38$} & \multicolumn{1}{r}{$0.06$} & \multicolumn{1}{r}{$0.27$}& \multicolumn{1}{r}{$0.49$}& $1.000$\\			
           & \multicolumn{1}{c}{$\delta_C$}     & \multicolumn{1}{r}{$0.55$} & \multicolumn{1}{r}{$0.06$} & \multicolumn{1}{r}{$0.43$}& \multicolumn{1}{r}{$0.68$} & $1.000$\\
		    & \multicolumn{1}{c}{$\sigma_{CH}$}   & \multicolumn{1}{r}{$0.07$} & \multicolumn{1}{r}{$0.05$} & \multicolumn{1}{r}{$0.00$}& \multicolumn{1}{r}{$0.18$} & $1.011$\\
			& \multicolumn{1}{c}{$\sigma_{HC}$}   & \multicolumn{1}{r}{$0.10$} & \multicolumn{1}{r}{$0.07$} & \multicolumn{1}{r}{$0.00$}& \multicolumn{1}{r}{$0.25$}& $1.048$\\ 
\noalign{\global\arrayrulewidth=0.1mm} \arrayrulecolor{manatee}\hline
   Observable & \multicolumn{1}{c}{$\alpha_{C}$}     & \multicolumn{1}{r}{$-0.15$} & \multicolumn{1}{r}{$0.07$} & \multicolumn{1}{r}{$-0.29$}& \multicolumn{1}{r}{$-0.01$}& $0.999$\\
			& \multicolumn{1}{c}{$\alpha_{H}$}     & \multicolumn{1}{r}{$12.59$} & \multicolumn{1}{r}{$1.19$} & \multicolumn{1}{r}{$10.52$}& \multicolumn{1}{r}{$14.97$}& $1.014$\\
			& \multicolumn{1}{c}{$\alpha_{d}$}     & \multicolumn{1}{r}{$-0.42$} & \multicolumn{1}{r}{$0.05$} & \multicolumn{1}{r}{$-0.51$}& \multicolumn{1}{r}{$-0.33$}& $1.017$\\
   		& \multicolumn{1}{c}{$\alpha_{FT}$}     & \multicolumn{1}{r}{$6.37$} & \multicolumn{1}{r}{$0.69$} & \multicolumn{1}{r}{$5.16$}&       \multicolumn{1}{r}{$7.75$}& $1.017$\\
			& \multicolumn{1}{c}{$\sigma_a$}   & \multicolumn{1}{r}{$0.15$} & \multicolumn{1}{r}{$0.08$} & \multicolumn{1}{r}{$0.00$}& \multicolumn{1}{r}{$0.31$}& $1.024$\\\noalign{\hrule height 1pt}
		\end{tabular}
	\end{table}

	The posterior distribution  provides useful information on the general performance of the Miami Heat team in a game. Table \ref{tab:post} shows the posterior summary for the parameters and hyperparameters included in the model. Furthermore, we computed a standard convergence diagnostic measure, $\hat{R}$ \citep{gelman2013bayesian}, for each parameter and hyperparameter. It is noteworthy that all $\hat{R}$ values were close to 1, which indicates good convergence of the MCMC algorithm.
	
\begin{figure}[H]
		\centering
		\subfigure[($p^{(CC)}$, $p^{(CH)}$)]{\includegraphics[width=62mm]{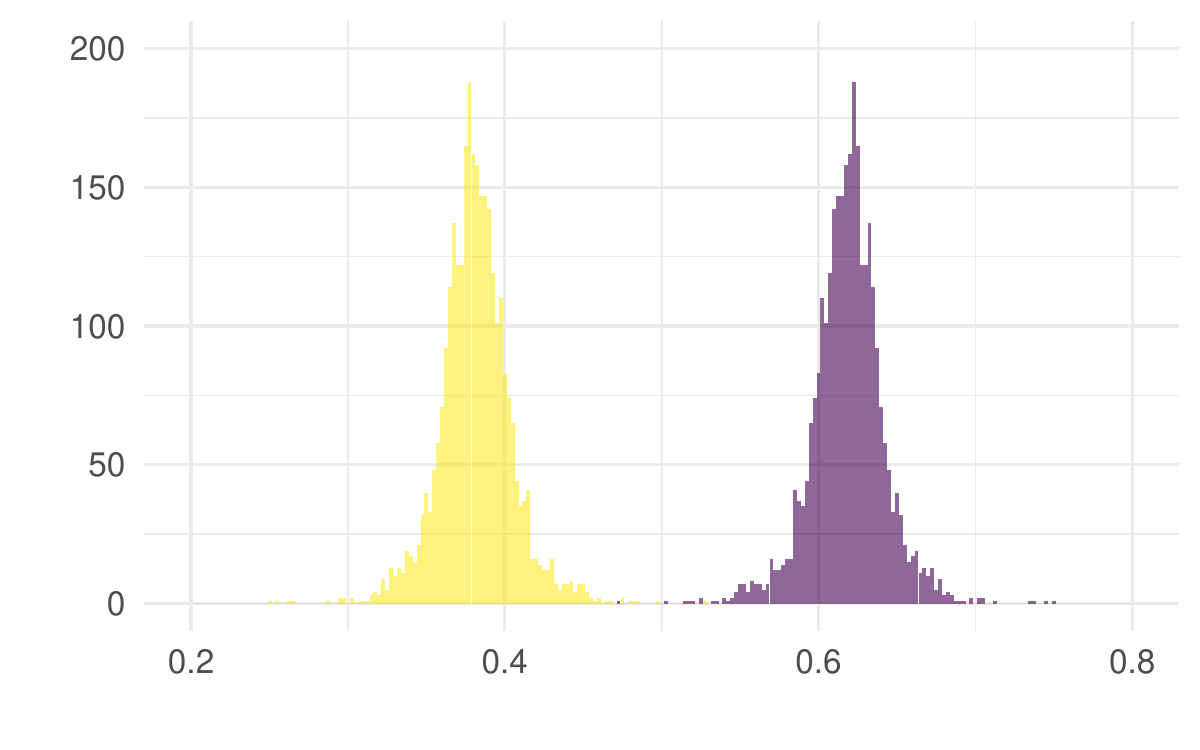}}
		\subfigure[($p^{(HC)}$, $p^{(HH)}$)]{\includegraphics[width=62mm]{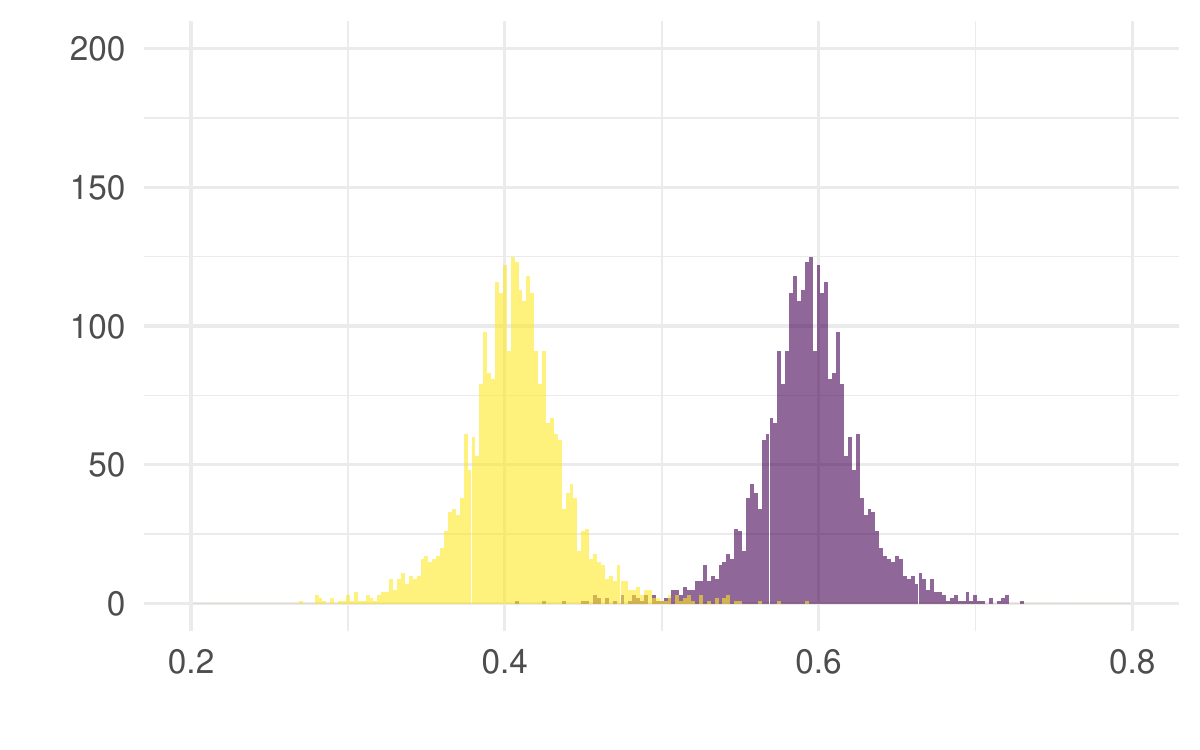}}
		\caption{Posterior distribution histograms (with frequencies) of the transition probabilities (a)  $p^{(CC)}$ (in purple) and   $p^{(CH)}$ (in yellow), and (b) $p^{(HC)}$ (in purple) and   $p^{(HH)}$ (in yellow) integrating the random effects associated with the game.}
		\label{fig:trans_prob}
	\end{figure}

	We focus first on the elements of the hidden model.  Posterior means of the common intercepts included in the transition probabilities $\beta_{CH}$ and $\beta_{HC}$ clearly indicate a higher probability of moving from the {\it hot }state to the {\it cold }state than from the {\it cold }state to the {\it hot }state.  In particular,    the posterior mean (computed by Formula \ref{eq:int_re}) of the probability of switching from $C$ to $H$ in a generic game is $0.38$, and from $H$ to $C$ is $0.59$. Thus, remaining in the {\it cold }state in one transition is more likely than switching to the hot, or than remaining in the {\it hot }state. This can also be  observed in  Figure \ref{fig:trans_prob}, which displays the posterior distribution of the transition probabilities for a generic game. Moreover, the expected posterior probability of being in the {\it cold }state when the game starts is around $0.55$ according to the posterior mean of $\delta_C$.
 
 On the other hand, for the observed Bernoulli sub-process the posterior distribution of the common intercepts $\alpha_{C}$ and $\alpha_{H}$ shows  a large difference of the magnitude of $\alpha_{H}$ with respect to $\alpha_{C}$  as seen in their means 
    E$(\alpha_{C} \mid \mathcal D)= -0.15$ and   E$(\alpha_{H} \mid \mathcal D)=$12.59. Following the same procedure as above for the transition probabilities, we have a posterior expected probability $0.46$ of making a shot from a distance of 0 feet to the basket  when the team is in the {\it cold }state, and almost $1$ in the case of the {\it hot }state when the distance to the basket is 0.25 feet.

	The two covariates considered in the modelling are relevant. The coefficient parameter associated with the distance to the basket, $\alpha_d$, is completely negative, E$(\alpha_d \mid \mathcal D)=$-0.42, with a small posterior standard deviation, SD$(\alpha_d \mid \mathcal D)$=0.03, which means that the probability of successful shooting decreases when the distance is further away. In addition, the credible interval of the coefficient associated with the {\it free throws} has a large positive expected value, E$(\alpha_{FT} \mid \mathcal D)=$6.37. Therefore, making a {\it free throw} looks easier than making a {\it field goal} from that distance. Finally, the random effects associated with the games included in the success probability  are also important since their associated standard deviation $\sigma_a$ has a posterior mean $0.15$. It can be stated that a relevant portion of the variability can be attributed to the variations in the scoring efficiency of the team in the different games.

All information we extract from the posterior distribution provides an immense number of possibilities   to assess different aspects of the performance of the team. We can discuss some of them right now.

	\subsubsection{Occupancy times}
\begin{figure}[H]
		\centering
		\subfigure[]{\includegraphics[width=62mm]{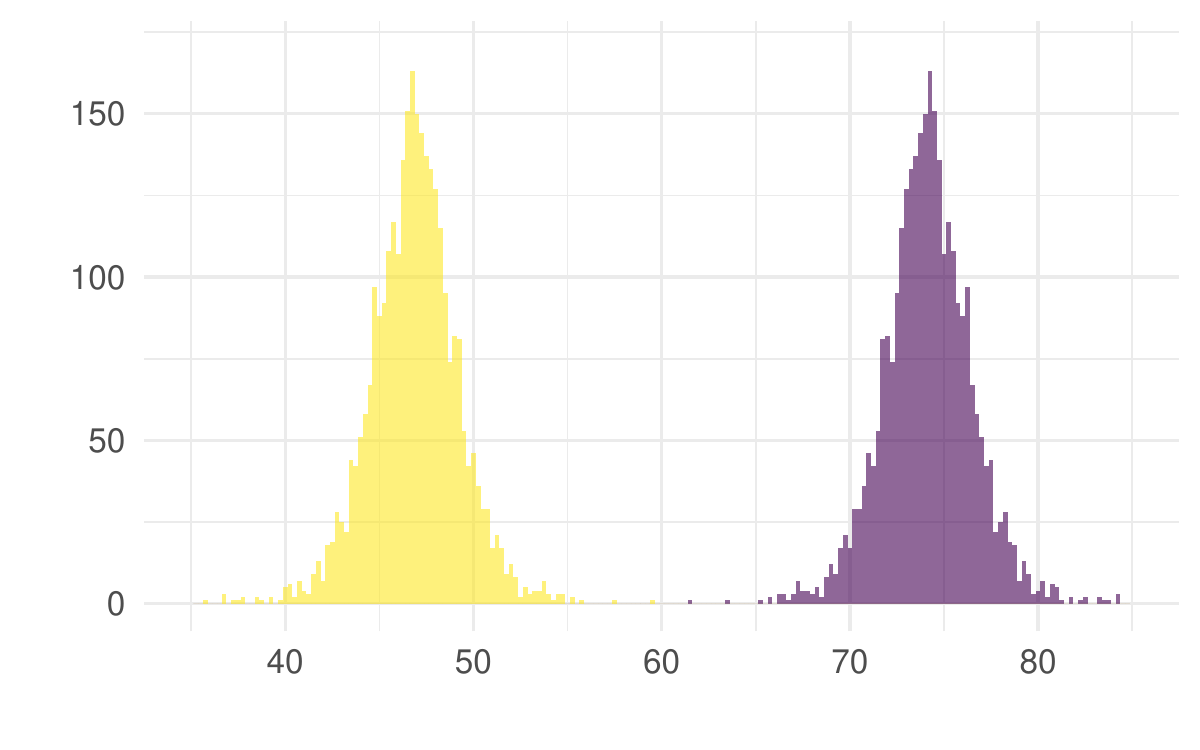}}
		\subfigure[]{\includegraphics[width=62mm]{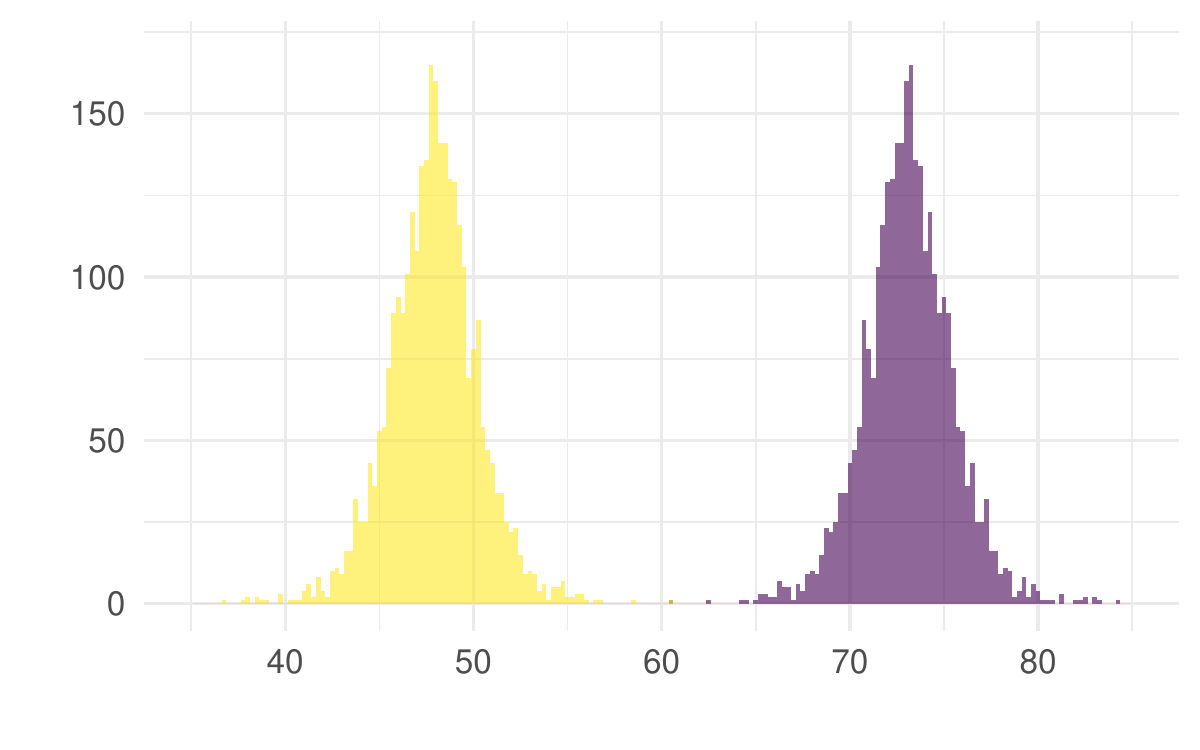}}
		\caption{Posterior distribution histograms (with frequencies) of the occupancy times in the states {\it hot }(in yellow) and {\it cold }(in purple)  in a game with $N=120$ shots that starts (a) in the {\it cold }and (b) in the {\it hot }state.}
		\label{fig:oc}
	\end{figure}

    Occupancy times admit many different posterior outputs. Here, for illustration we only focus  on the posterior distribution of occupancy times in   the {\it cold }and the {\it hot }state in a generic game with $N=120$ shots.
     Figure \ref{fig:oc} shows the posterior distribution of the occupancy times in the {\it cold }and the {\it hot }state  and   the two possibilities of initial state, {\it cold }and {\it hot}.

    It is interesting to note that the team spends more time in the {\it cold }state than in the {\it hot } (with a posterior mean of 74.05 and 46.85, respectively),
 and that the state in which the team starts playing is practically irrelevant because there is hardly any difference between the two figures. 
 
\subsubsection{Stationary distributions}

 Although we could consider a stationary distribution of the chain associated with each game of the season, we focus on the posterior distribution  $\pi((\Delta^{(C)}, \Delta^{(H)} \mid \boldsymbol \theta) \mid \mathcal D)$ of the stationary distribution of the chain corresponding to a generic game. Table \ref{tab:post_st} shows a summary of  the  posterior distribution  of the stationary distribution  for the {\it cold }and {\it hot }state. 
 \begin{table}[H]
		\centering
		\caption{Posterior summaries (mean, standard deviation and 95\% credible interval) for the stationary distribution of the hidden Markov chain associated with a generic game.}\label{tab:post_st}
		\vspace{0.25cm}
		\begin{tabular}{ccccc}
			\noalign{\hrule height 1pt}
			\multicolumn{1}{l}{} & \multicolumn{1}{c}{mean}    & \multicolumn{1}{c}{sd} & $q_{0.025}$ & $q_{0.975}$ \\ \noalign{\hrule height 1pt}
			\multicolumn{1}{c}{$\Delta^{(C)}$}  & \multicolumn{1}{r}{$0.61$} & \multicolumn{1}{r}{$0.02$} & \multicolumn{1}{r}{$0.57$} & \multicolumn{1}{r}{$0.65$}\\ 
            \multicolumn{1}{c}{$\Delta^{(H)}$}  & \multicolumn{1}{r}{$0.39$} & \multicolumn{1}{r}{$0.02$} & \multicolumn{1}{r}{$0.35$} & \multicolumn{1}{r}{$0.43$}\\ \noalign{\hrule height 1pt}
		\end{tabular}
	\end{table}
	
	In summary, the probability of being in the {\it cold }state for the Miami Heat team in the 2005-2006 season was around $0.61$ (thus, the probability of being in the {\it hot }state it was $0.39$). Both distributions have very little variability.

 \subsubsection{Sojourn times}
    
\begin{figure}[ht]
\begin{center}
\includegraphics[width= 8cm]{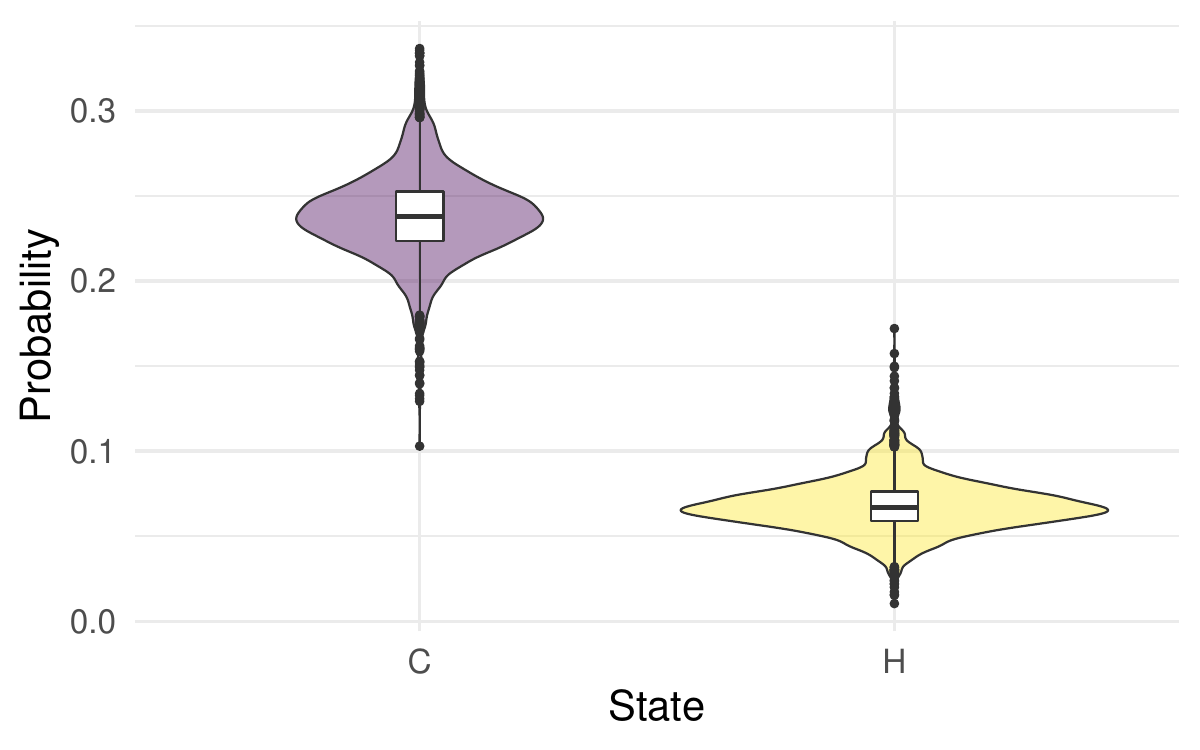} 
\caption{Violin and box plot of the posterior probability distribution for a {\it cold }and  a {\it hot }streak.} \label{fig:rem_times}
\end{center}
\end{figure}

    We consider a {\it cold }or a {\it hot }streak when the team stays in the same state for more than three shots. This number of shots is an arbitrary choice that can only be justified in order to illustrate the potential of our modelling and improve the understanding of the behaviour of the team.  Figure \ref{fig:rem_times} shows 
    a violin plot of the posterior distribution of the probability for a {\it cold }and for a {\it hot }streak in a generic game of the Miami Heat. There, we observe the probability of a {\it cold }streak (around 0.25) is nearly three times higher than the probability of a {\it hot }streak (less than 0.1). The much smaller amplitude of the {\it hot }distribution compared to the {\it cold }one is also evident.

    \subsubsection{Probability of making a basket}
    
\begin{figure}[H]
\begin{center}
\includegraphics[width= 11cm]{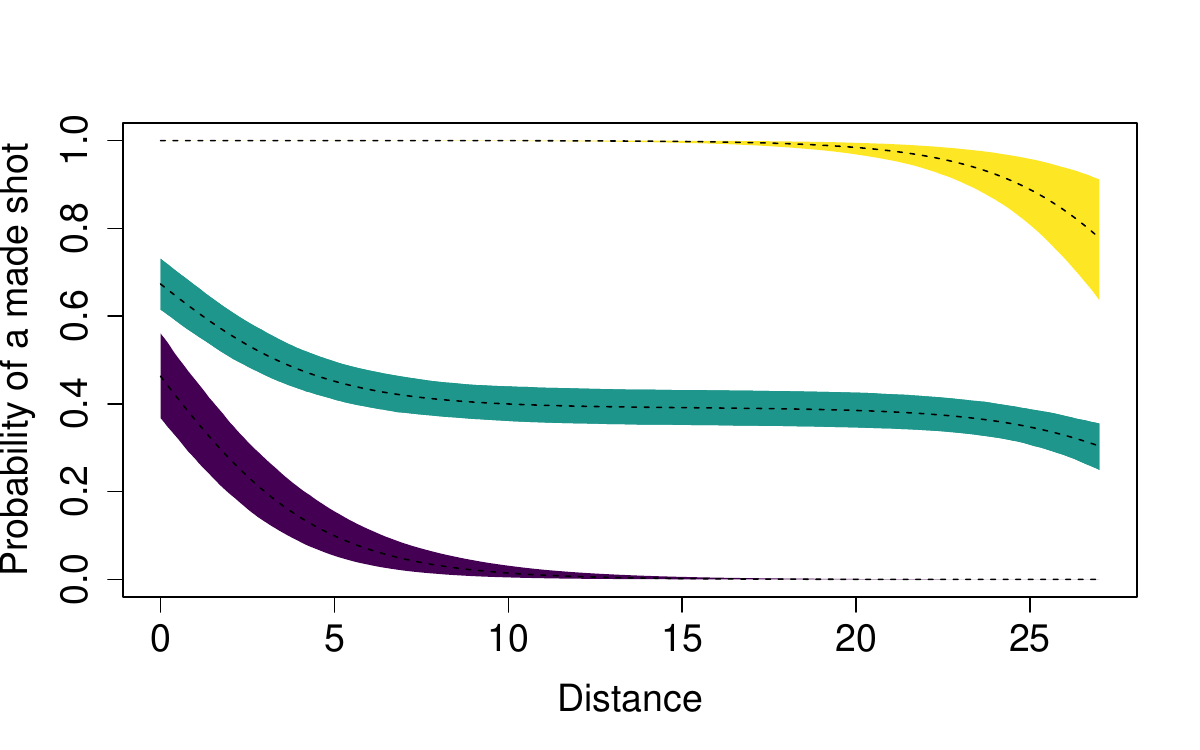} 
\caption{Posterior means  and 95$\%$ credible interval for the probabilities of making a basket depending on the distance, in feet,  in three different scenarios. Purple curve represents the case when the team is in the {\it cold }state, the yellow plot   is for the {\it hot }state, and the green plot is the general case when the state of the team is unknown.} \label{fig:succes_distance}
\end{center}
\end{figure}

The probability of making a basket depends on the state the team is currently visiting, the distance to the basket at which the shot is made, and whether the shot is a {\it free throw} or not. Figure \ref{fig:succes_distance} shows the posterior  mean and the 95\% credible interval for the probability of making a basket depending on the distance at which the shot (non-{\it free throw}) is made when the team is in the {\it cold }state, {\it hot }state or in the case where the state of the team is unknown. 

When the team is in the {\it cold }state, one can see that for easy shots (i.e. shots close to the basket) there is a probability of around $0.5$ of making a basket. However, from a distance of 10 feet, it is almost impossible to succeed.
On the other hand,  when the team is in the {\it hot }state, it is very likely to make a successful shot up to 15 feet. Then, from this distance, the probability starts to  decrease slowly. Further, for a shot, in which the current state is unknown, the probability of success is also negatively related to the distance to the basket. Shots closer to the basket have a probability of success around $0.7$, whereas for intermediate shots this probability stabilises at around $0.5$. Finally, the probability of making a three-point shot drops to $0.4$.

        \begin{figure}[H]
\begin{center}
\includegraphics[width= 11cm]{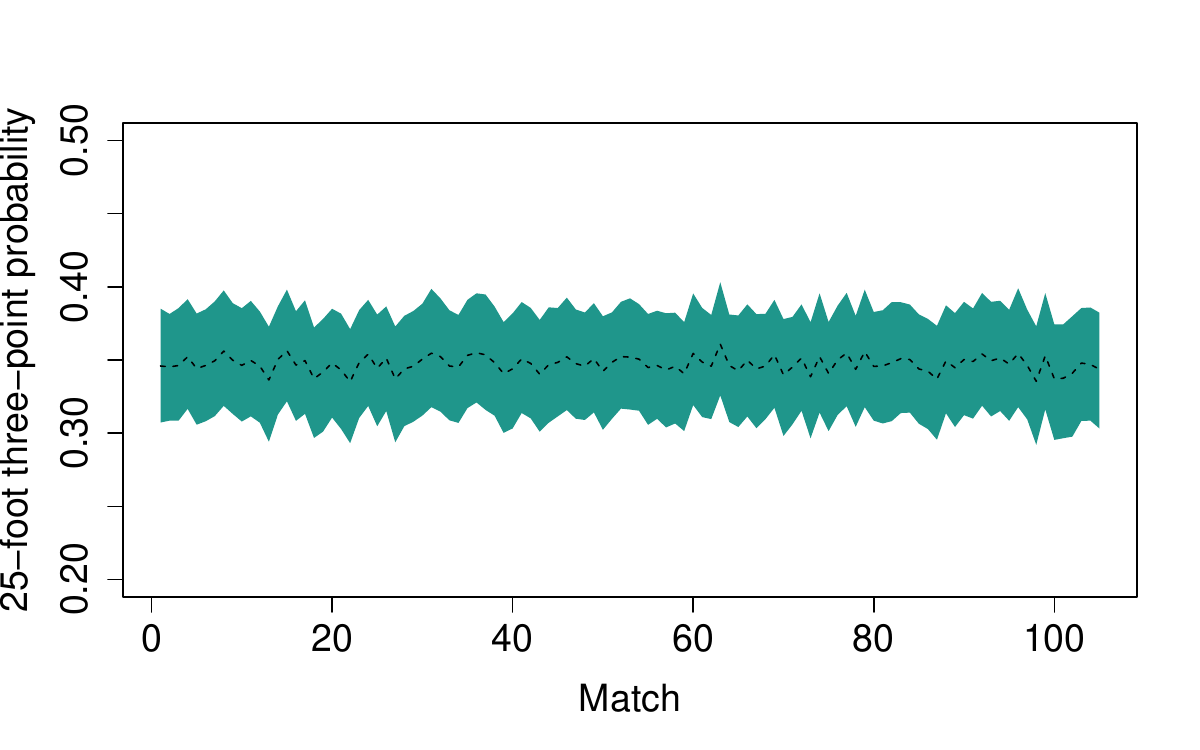} 
\caption{Posterior mean and credible interval for the probability of making a basket from 25 feet distance in the general case when the state of the team is unknown.} \label{fig:succes_25}
\end{center}
\end{figure}

It can be interesting to visualise the team's performance over the different games of the season. In that sense, Figure \ref{fig:succes_25} shows the mean and the 95th percentile credible interval of the probability of hitting a shot from 25 feet, in which the team’s state is unknown, in each of the games of the season. We observe a very marked regularity in all of the games of the season,   very little variability among games and a very stable behaviour in all phases of the season. 

{\subsection{Evidence in favour of the BLHMM} }

Now, we compare the previous model, denoted as $\mathcal{M}_1$, with a simpler model, $\mathcal{M}_2$, without hidden structure and joint distribution factorised as:
\begin{equation*}
  f(\boldsymbol{Y}, \boldsymbol{\theta}, \boldsymbol{\psi}) =   \big(\prod_{i=1}^{N} \prod_{n=1}^{M_i} \, f({Y}_{in} \mid  \boldsymbol{\theta}, \boldsymbol{\psi})    \ 
   \big)\  f(\boldsymbol{\psi}|\boldsymbol \theta) \ \pi(\boldsymbol{\theta}). 
\end{equation*}

\noindent Here, each variable ${Y}_{in}$ follows a Bernoulli distribution with a probability parameter $\gamma_{in}$:

\begin{equation*}
    (Y_{in}|\boldsymbol{\theta},  \boldsymbol \psi) \sim \text{Bern}(\gamma_{in}).
\end{equation*}

\noindent Each success probability parameter $\gamma_{in}$ is associated with the covariates and the random effects through the logit function, as follows:
\begin{align*}
\text{logit}(\gamma_{in}) = \alpha + \alpha_{d}X_{in} + \alpha_{FT} I_{FT}(in) + a_i.
\end{align*}
\noindent In this expression, all elements of the model have  similar meaning as we indicated for model $\mathcal{M}_1$. In addition, since this is a nested model of $\mathcal{M}_1$, we chose the same prior distribution for all parameters and hyperparameters present in $\mathcal{M}_2$.

In order to conduct a model comparison between $\mathcal{M}_1$ and $\mathcal{M}_2$, we compute the conditional predictive ordinate ($CPO$) for each observation from the cross-validated predictive density, as in \citet{gelfand1994}. The fundamental idea underlying this approach is based on the assumption that if the estimated model is correct, each observation can be considered as a random variable drawn from the cross-validated predictive density. The $CPO$s for models $\mathcal{M}_1$ and $\mathcal{M}_2$ are defined as
\begin{align}
   CPO_{in}^{\mathcal{M}_1} &= f( {Y_{in}} \mid \mathcal D^{-(in)},\mathcal{M}_1), \\
        CPO_{in}^{\mathcal{M}_2} &= f( {Y_{in}} \mid \mathcal D^{-(in)},\mathcal{M}_2).
\end{align}
\noindent Here, $\mathcal D^{-(in)}$ denotes all of the data in $\mathcal D$ except for the $n$-th shot in game $i$. Higher $CPO$s values from a particular model compared to other provide support for that model. We compute these values following the approach in \citet{Ntzoufras}. This procedure applies the self-normalized importance sampling technique  to  approximate $CPO$ values obtained from samples of the posterior distribution computed through the complete data $\mathcal D$.

Specifically, we consider the difference $CPO_{d_{in}} = CPO_{in}^{\mathcal{M}_1} - CPO_{in}^{\mathcal{M}_2}$ as a normally distributed random variable, i. e.,

\begin{equation}
    (CPO_{d_{in}} \mid \mu_d, \sigma_d ) \sim \text{N}(\mu_d, \sigma_d^2).
\end{equation}

\noindent In this context, bolstered by the central limit theorem, we chose a broad normal distribution for the marginal prior of the mean parameter, i.e., $\pi(\mu_d)=\text{N}(0,10^2)$. In addition, we opted for a uniform distribution for the standard deviation, $\pi(\sigma_d)=\text{U}(0,10)$.

\begin{figure}[ht]
\begin{center}
\includegraphics[width= 11cm]{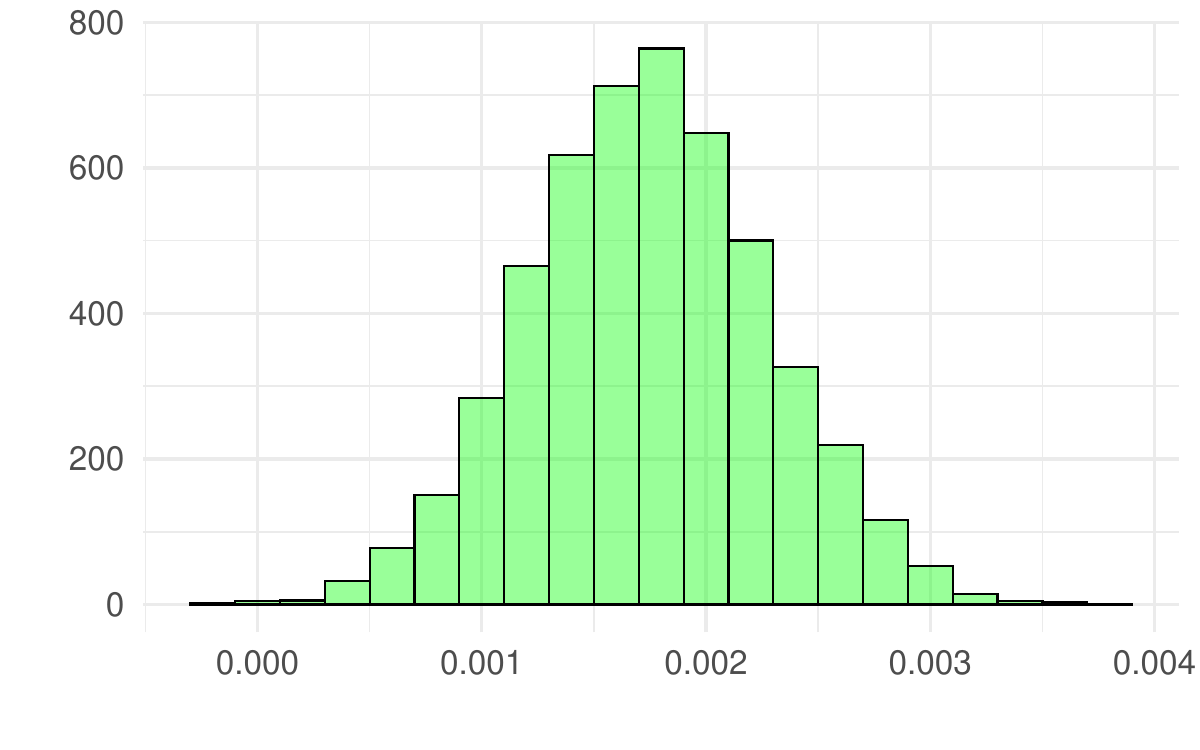} 
\caption{Posterior distribution histogram of the mean parameter $\mu_d$.} \label{fig:mud}
\end{center}
\end{figure}

Figure \ref{fig:mud} shows the approximate posterior distribution of $\mu_d$. It is evident that the marginal posterior is almost entirely positive, providing substantial evidence in favour of $\mathcal{M}_1$ over $\mathcal{M}_2$. In other words, the BLHMM provides a better fit for this data than a model lacking hidden structure.

\section{Conclusions}

During our analysis of the Miami Heat season, we noticed the {\it cold hand} more likely than the {\it hot hand}. In fact, based on our results, in any given game, the {\it cold streaks} seem to have more occurrences than {\it hot streaks}. Moreover, according to the posterior distribution of the transition probabilities, the two states are clearly differentiated: in fact, the team's performance is very different depending on which state it is in.

We believe our BLHMM can be employed as a valuable tool for analysing the performance of a team during a match. Moreover, it could 
be used to compare the performances of different teams, providing information that is not possible to obtain through the traditional modelling techniques. {It is worth noting that we provide evidence that supports our approach over a simpler model.} Additionally, it has the potential to be applied in the analysis of an opponent's team performance during a game.

Finally, for future research, the inclusion of a third intermediate state that is neither \textit{hot} nor \textit{cold} is an interesting possibility.  It could also be highly valuable to incorporate the match time as a covariate in the transition matrix. By doing so, we would be able to study how the team behaves at each specific moment of the match. In this sense, exploring continuous-time Markov chains could be a worthwhile approach to better model this phenomenon.

\section*{Acknowledgements}
Gabriel Calvo's research was partially funded by the ONCE Foundation, the Universia Foundation, and the Spanish Ministry of Education and Professional Training, grant FPU18/03101. Carmen Armero and Gabriel Calvo's research was partially funded by the Spanish Research project Bayes$\_$COCO (PID2019-106341GB-I00) from the Ministry of Science and Innovation Grant. Luigi Spezia's research was funded by the Scottish Government's Rural and Environment Science and Analytical Services Division. Comments from Fergus Chadwick improved the quality of the final paper.

\bibliographystyle{apalike}
\bibliography{hot_hand_bib}

\end{document}